\documentclass[twocolumn,showpacs,aps,prl]{revtex4-1}

\usepackage{bm}
\usepackage{amssymb,xcolor}
\usepackage{dcolumn}
\usepackage{bm}
\usepackage{graphicx}
\usepackage{amsmath}
\usepackage{latexsym}
\usepackage{amsfonts}
\usepackage{amssymb}
\usepackage{array}
\usepackage{epsfig}
\usepackage{epstopdf}
\usepackage{times}
\usepackage{amsmath,epsfig}

\newcommand{\beq}{\begin{equation}}
\newcommand{\eeq}{\end{equation}}

\newcommand{\bra}[1]{\langle#1|}

\newcommand{\ket}[1]{|#1\rangle}

\newcommand{\id}{\leavevmode\hbox{\small1\normalsize\kern-.33em1}}

\begin{document}

\title{Experimental achievement of the entanglement assisted capacity for the 
depolarizing channel}

\author{Andrea Chiuri$^{1}$, Sandro Giacomini$^{1}$, Chiara Macchiavello$^{2}$, and Paolo Mataloni$^{1,3}$}
\affiliation{$^1$ Dipartimento di Fisica, Sapienza Universit\`a di Roma, Piazzale Aldo Moro 5, I-00185 Roma, Italy\\
$^2$ Dipartimento di Fisica and INFN-Sezione di Pavia, via Bassi 6, 27100 Pavia, Italy\\
$^3$ Istituto Nazionale di Ottica (INO-CNR), Largo E. Fermi 6, I-50125 Firenze, Italy
}

\date{\today}
\begin{abstract}
We experimentally demonstrate the achievement of the entanglement assisted 
capacity for classical information transmission over a depolarizing channel.
The implementation is based on the generation and local manipulation of 
2-qubit Bell states, which are finally measured at the receiver by a complete 
Bell state analysis. The depolarizing channel is realized by introducing 
quantum noise in a controlled way on one of the two qubits. 
This work demonstrates the achievement of the maximum allowed amount of information 
that can be shared in the presence of noise and the highest reported value in the noiseless case.
\end{abstract}


\maketitle

{\it Introduction.- }
Noise is unavoidably present in any realistic implementation of a
communication channel. It is therefore of great importance to design 
strategies that 
allow to optimise the flow of information transmitted in presence of noise. 
In the most general scenario information can be transmitted by quantum states
and several notions of efficiency can be defined according to the considered 
task and the 
resource available along the transmission channel. For example, a quantum 
communication channel can be designed to transmit classical \cite{class}, 
private classical \cite{priv-class} or 
quantum information \cite{q-info}, 
and it can be employed on its own or with the addition 
of other resources, such as entanglement.
The case of an entanglement assisted quantum communication channel 
occurs when classical information is transmitted and entanglement is a 
priori available between sender and receiver \cite{bennett}.
The capacity is then given by the maximum amount of information that can be 
transmitted over the channel by optimising the input signals and the output 
decoding procedure.
In this paper we consider the latter scenario and we present an experimental 
demonstration of the best performance obtained by an 
entanglement assisted depolarizing channel for qubits, that allows to
achieve the information capacity. 
The experimental realization presented in this work relies on a quantum 
optical implementation, but it lays the ground for new perspectives of 
applications to a great variety of communication scenarios.
We want to stress that this is the first experimental demonstration of the
information capacity for a controlled noisy quantum communication channel, 
since previous experimental realisations consider just the case of a 
noiseless channel \cite{kwiat} or classical noise \cite{prev11prl}.

An entanglement assisted communication scenario is given by a 
(generally noisy) quantum channel along which quantum states can be 
transmitted 
by assuming that an unlimited amount of noiseless entanglement is a priori 
available between the sender and the receiver \cite{bennett}.
The entanglement assisted classical capacity (EACC) is then given by the 
maximum amount of
classical mutual information that can be transmitted over such a channel.
In this work we consider the case of a depolarizing qubit channel 
\cite{nielsen}, where the action on a given quantum state of a 
two-dimensional system $\rho$ can be described as

\begin{equation}\label{channel}
\Gamma_{\{p\}}[\rho]= {\sum^3}_{i=0}{p_i \sigma_i \rho \sigma_i}
\end{equation}
Here $\sigma_0$ is the identity operator, \{$\sigma_i$\} ($i=1,2,3$) 
are the three Pauli operators $\sigma_x ,\sigma_y, \sigma_z$ respectively, and
$p_0=1-p$ (with $p\in[0,1]$), while $p_i=p/3$ for $i=1,2,3$.

In this case the entanglement assisted classical capacity takes the simple 
form
\cite{bennett}

\begin{equation}\label{midep}
C = 2 + (1-p)\log_2 (1-p) + p \log_2 (p/3)\;.
\end{equation}

The scheme considered in this work is composed of two qubits, 
initially prepared in the singlet state $\ket{\psi^-}=
\frac{1}{\sqrt 2}(\ket{01}-\ket{10})$, where the states $\ket{0}$ and 
$\ket{1}$ are a basis for each qubit. The singlet state then represents
the noiseless entangled state that is a priori shared by the sender and the 
receiver. Classical information is then encoded by the sender by performing,
with equal probabilities, either the identity or one of the three Pauli 
operators on his qubit, which is then transmitted along the depolarizing 
channel to the receiver. The receiver finally performs a Bell measurement 
in order to retrieve the information encoded in the two-qubit system. 
This scenario allows to achieve the capacity (\ref{midep})
\cite{shad10njp}.

\begin{figure*}[t!!]
\includegraphics[width=\textwidth]{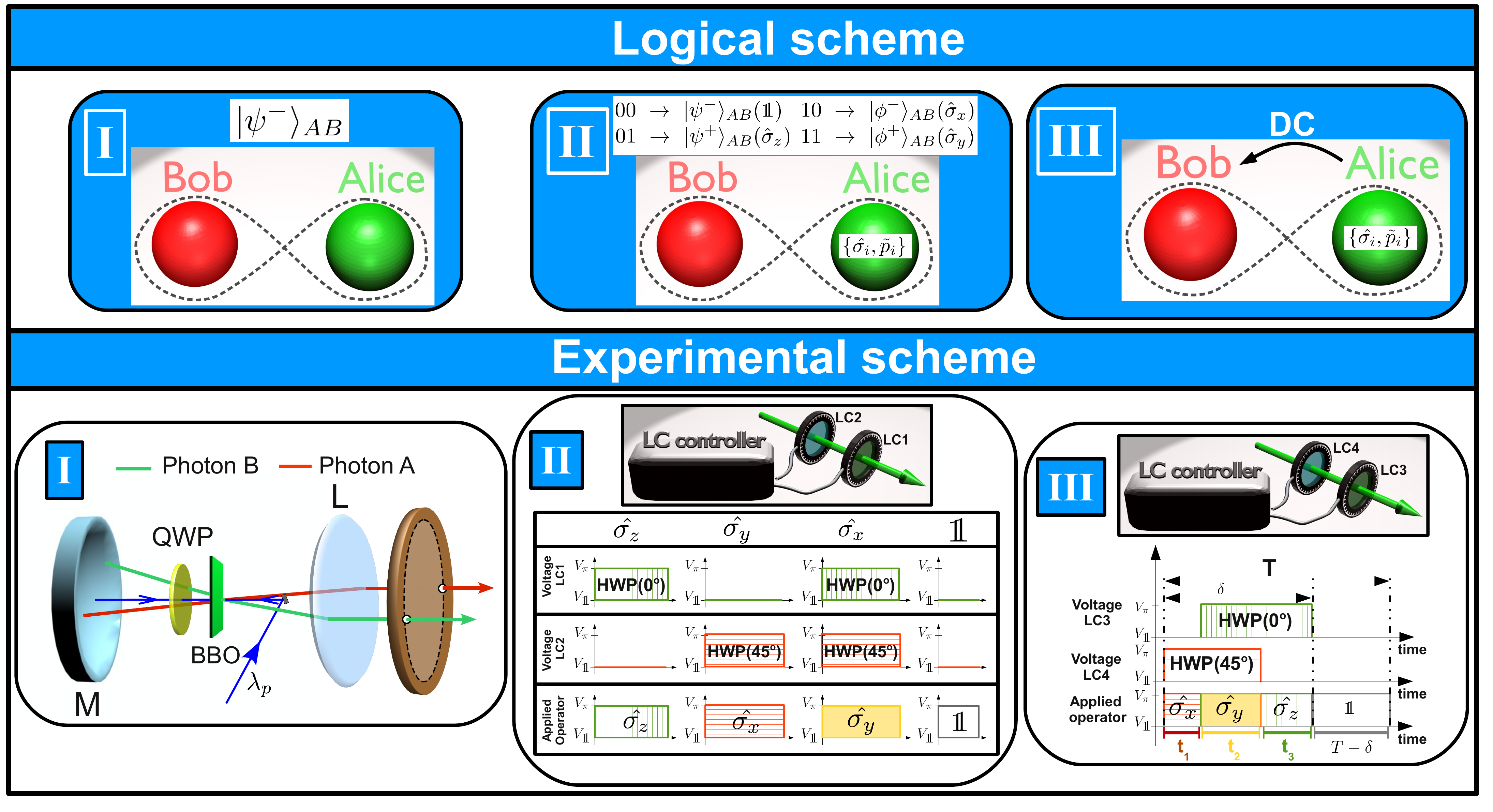}
\caption{I) Spontaneous parametric down conversion (SPDC) source of 
polarization entanglement. 
A Beta-Barium Borate (BBO) nonlinear crystal 
is shined by a vertically polarized continuous
wave (cw) UV-laser ($\lambda_p$ = 364 nm), and the two
photons are emitted at degenerate wavelength $\lambda = 728$ nm with horizontal polarization.
Polarization entanglement is generated by the double passage
(back and forth, after the reflection on the spherical mirror M) of
the UV beam. The backward emission generates the so-called
V cone: the SPDC horizontally polarized photons passing twice
through the quarter-wave plate (QWP) are transformed into
vertically polarized photons. The forward emission generates the
H cone. 
Because of the temporal and spatial superposition, the
indistinguishability of the two perpendicularly polarized SPDC
cones creates polarization entanglement $({\ket{H}}_A{\ket{H}}_B+{\ket{V}}_A{\ket{V}}_B)/\sqrt{2}$.
The relative phase between the states $\ket{HH}_{AB}$ and $\ket{VV}_{AB}$ can be varied by
translation of spherical mirror M. A lens $L$ located at a focal
distance from the crystal transforms the conical emission into
a cylindrical one. II) Alice encodes two bits of classical information by applying 
single qubit Pauli transformations on one entangled particle. This can be realized by applying a 
suitable voltage to the liquid crystal modulator, LC1 and LC2 acting as waveplates and implementing the transformations 
$\{\hat{\sigma}_i, \tilde{p}_i\}$. 
III) Photon A is transmitted through a depolarizing channel (DC), here implemented by liquid crystal modulators LC3 and LC4. 
The LCs's activation time corresponding to the three Pauli operators 
$\hat{\sigma}_x$, $\hat{\sigma}_y$, $\hat{\sigma}_z$ is the same, i.e. $t_1=t_2=t_3$.}
\label{schema}
\end{figure*}

In this work we implement an experimental scheme corresponding to the above
scenario and demonstrate that the capacity (\ref{midep}) can actually be 
achieved, allowing the optimal transmission of classical information through
the channel and therefore the use of the channel at the best of its possible 
performances.
We then measure the classical information transmitted through the channel in 
terms 
of the classical mutual information \cite{classinfo}, which can be expressed
as

\begin{equation}\label{mi}
I = \sum_{x} p_1(x)\sum_{y} p(y|x) \log_2 \frac{p(y|x)}{p_2(y)} 
\end{equation}

where $x$ and $y$ are the input/output variables, with corresponding 
probability distributions $p_1(x)$ and $p_2(y)$, while $p(y|x)$ represents 
the conditional probability of receiving $y$ given transmission of $x$.
In our experiment the variables $x$ and $y$ correspond to the four Bell 
states, $p_1(x)=1/4$ and $p(y|x)$ is the conditional probability of 
detecting the Bell state $y$ given that the Bell state $x$ is transmitted. 


{\it General scheme.- }
We describe here the general experimental scheme to implementing 
the scenario described above.
The two-photon Bell states are engineered by exploiting 
a polarization entanglement source \cite{cine04pra,barb05pra} 
[See Fig.\ref{schema}I)] which allows to generate accurately the states 
$\ket{\phi^{\pm}}=\frac{1}{\sqrt{2}}(\ket{00} \pm \ket{11})$,
with qubit $\ket{0}$ ($\ket{1}$) corresponding to the horizontal $H$ 
(vertical $V$) polarization of the photon. 
The other Bell states $\ket{\psi^{\pm}}=\frac{1}{\sqrt{2}}(\ket{01} \pm \ket{10})$ 
can be obtained by applying simple single-qubit local operations. 
Let us now assume that the initial state, shared by the sender Alice and the 
receiver Bob, is represented by the singlet 
$\ket{\psi^{-}}_{AB}=\frac{1}{\sqrt{2}}(\ket{01} - \ket{10})$.
In this case Alice encodes classical information by performing local 
transformations on 
photon A, which is then transmitted to Bob through the depolarising channel, 
while photon B belongs to Bob.

In the experiment the information encoding was performed by employing two 
liquid crystals (LC1 and LC2) acting 
simultaneously on photon A. This corresponds to
apply the single qubit local operation $\hat{\sigma}_{i}$. 
The LCs acted as phase retarders, with relative phase between the ordinary and 
extraordinary
radiation components depending on the applied voltage $V$.
Precisely, $V_{\pi}$ and $V_{\id}$ (Fig.\ref{schema}II) corresponded to the
case of LCs operating as half-waveplate (HWP) and as the
identity operator, respectively.
The LC1 and LC2 optical axes were set at $0^\circ$ and $45^\circ$ with respect 
to the V-polarization.
When the voltage $V_{\pi}$ was applied, LC1 (LC2) acted as a $\sigma_z$ ($\sigma_x$) on the single qubit.
The simultaneous application of $V_{\pi}$ on both LC1 and LC2 corresponds to the $\sigma_y $ operation.
As described in Fig.\ref{schema}II), LCs were suitably activated by a remote control in order to perform the four 
Pauli operators with equal probabilities. In this way, depending on the transformation implemented by the 
LCs, Alice encodes two bits of classical information in the shared 
Bell state, i.e. 
$ \id \rightarrow 00$, $ \hat{\sigma}_z \rightarrow 01$, 
$ \hat{\sigma}_x \rightarrow 10$, $ \hat{\sigma}_y \rightarrow 11$.

\begin{figure}[h!!]
\includegraphics[width=\columnwidth]{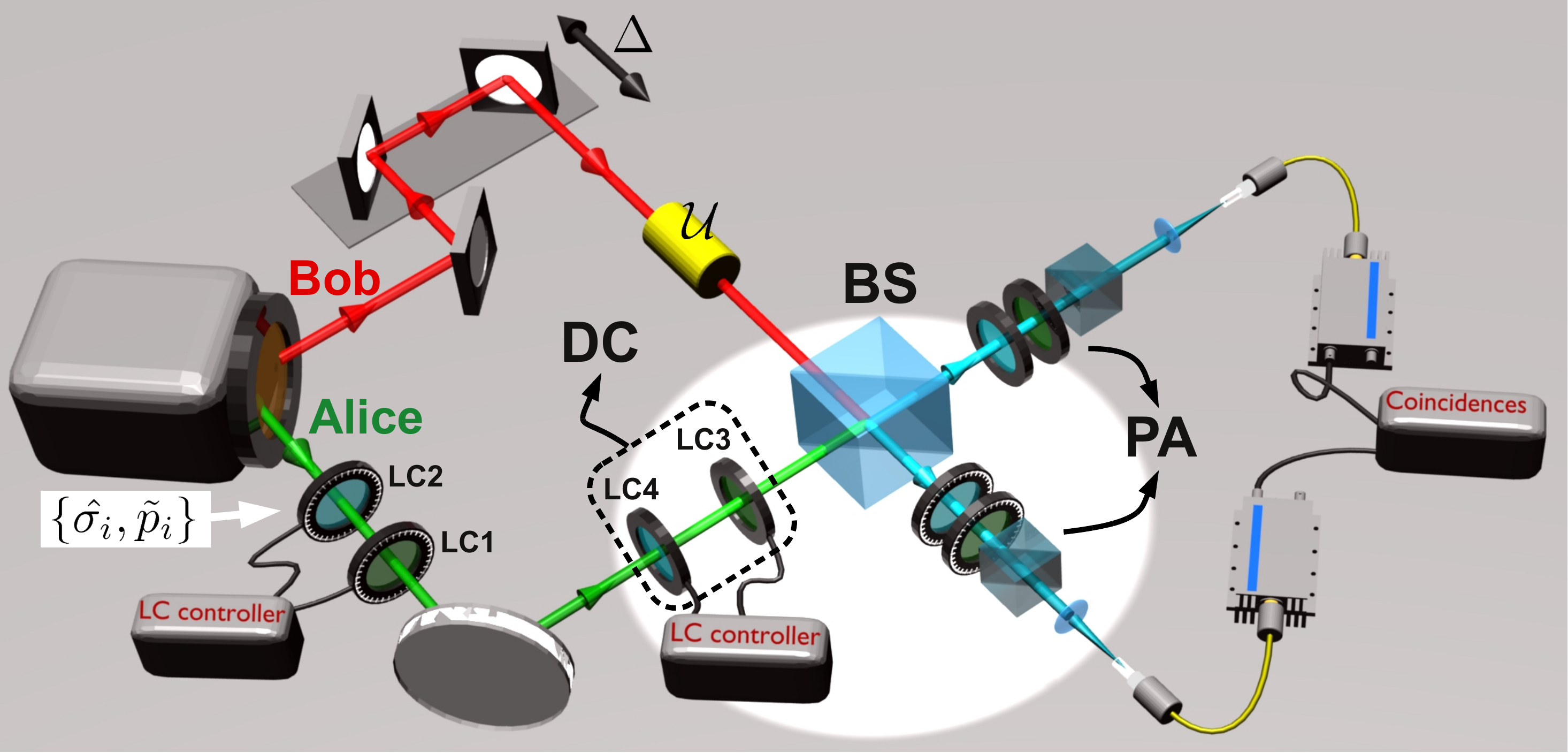}
\caption{Experimental setup. Alice and Bob share a two-qubit singlet state. 
LC1 and LC2 implement the four 
Pauli operators $\sigma_i$ with probability $\tilde{p}_i$. 
The LC3 and LC4 systems perform the DC operation. 
By properly setting both temporal delay and spatial mode superposition,  
the two photons arrive in the BS in condition of complete 
undistinguishability. 
The noisy state is projected onto the four Bell states by exploiting 
the BS and the transformation $\mathcal{U}$. Each polarization analysis (PA)
setup consists of a half-waveplate, a quarter-waveplate and a polarizing BS placed before the detectors.}
\label{setup}
\end{figure}

The manipulated qubit was then sent through a depolarizing channel (DC). 
Several experimental techniques have been previously proposed and realized 
to engineer a 
DC \cite{ricci04prl,karp08josa,saha11pra,jeonQP}.
In our case the noisy channel was implemented according to the scheme
proposed in \cite{chiu11prl} [See Fig.\ref{schema}III)]. 
The necessary logical operations were implemented 
by careful adjustment of the voltage applied to two
LCs and of the activation interval of each LC system.
In the next paragraph we will discuss in more detail how the realized noisy 
channel works.

In order to evaluate the joint probabilities, the four possible outputs $y$ 
need to be measured for each input state $x$. This final Bell measurement, corresponding 
to the projective measurement 
$\{ \ket{\psi^-}\bra{\psi^-}, \ket{\psi^+}\bra{\psi^+}, \ket{\phi^-}\bra{\phi^-}, \ket{\phi^+}\bra{\phi^+}\}$, 
can be realized by using a suitable interferometric setup and by 
performing those local transformations which allow to select 
the desired Bell state. 

{\it Experimental realization.- }
Let us consider the expression of the
classical mutual information given in Eq. (\ref{mi}). 
As mentioned above, after encoding, photon A is transmitted through the 
depolarizing channel. The experiment was performed by starting from the shared 
singlet state $\ket{\psi^-}$ and for each one of the other Bell states 
obtained after the encoding operation.
The DC was performed by two liquid crystal retarders 
(LC3 and LC4) inserted within the path of one of the two photons.
We were able to switch between $V_{\id}$ and $V_\pi$
in a controlled way and independently for both LC3 and LC4.
We could also adjust the temporal delay between the intervals corresponding to a 
$V_\pi$ voltage applied to the two retarders.
Let us define $t_1$, $t_2$, $t_3$ respectively as the activation time of the 
operators $\sigma_x$, $\sigma_y $ or $\sigma_z$ while $T$  is the period of 
the LCs activation cycle.
The condition $t_1=t_2=t_3$ corresponds to the case of a depolarizing channel, 
with the three Pauli operators acting on the single qubit with the same 
probability $p_{exp}=\frac{\delta}{T}=\frac{t_1+t_2+t_3}{T}$.
In the experiment, the parameter $p_{exp}$ was varyed by changing the interval 
$\delta$ for a fixed period $T$.

We show in Fig. \ref{setup} the actual interferometric setup adopted to 
perform projective measurements $\{ \ket{\psi^-}\bra{\psi^-}, 
\ket{\psi^+}\bra{\psi^+}, \ket{\phi^-}\bra{\phi^-}, 
\ket{\phi^+}\bra{\phi^+}\}$. 
The transformation $\mathcal{U}$, sketched in the same figure, represents the different
local unitaries necessary to transform each Bell state $\ket{\phi^+}$, $\ket{\phi^-}$, $\ket{\psi^+}$ into the
singlet state $\ket{\psi^-}$ ,
i.e. the only one that allows to measure coincidences between
the two output modes of the BS in condition of complete undistinguishability.
They are implemented by proper setting of the optical axes of a
quarter-wave plate and a half-wave plate.
In this way we could perform a complete Bell
measurement by performing the four projections at different times.

For each value of the noise degree, the probabilities associated to the measurement of 
each Bell state, corresponding to the conditional probabilities
$p(y|x)$ of Eq. (\ref{mi}), were obtained by considering the coincidence counts measured 
by the interferometer for each projection normalized over the results of the four measurements.
After the transmission through the beam splitter (BS), photons were coupled 
into single-mode fibers by using GRaded INdex 
(GRIN) lenses \cite{ross09prl} and detected by single photons detectors.   

{\it Experimental results.- }
Let us consider Eq. (\ref{mi}), with x,y=1,..,4 representing the four Bell 
states. In our case
$1\rightarrow \ket{\psi^-}$, $2\rightarrow \ket{\psi^+}$, $3\rightarrow \ket{\phi^-}$, $4\rightarrow \ket{\phi^+}$.
Since the probabilities of the four input Bell states are equal we have 
$p_1(x)=\frac{1}{4}$.
Therefore Eq. (\ref{mi}) reads

\begin{equation}\label{mi}
I_{meas} = \frac{1}{4}\sum_{x,y} p(y|x) \log_2 \frac{p(y|x)}{p_2(y)}\;. 
\end{equation}

The probabilities $p(y|x)$ in the above expression were measured by 
considering 
separately each Bell state $x$ and by projecting it 
onto the four Bell states after the action of the DC. 
The experiment was carried out for each 
state obtained after the encoding operation and for several values of the noise degrees, i.e. 
of the parameter $p_{exp}$. The values of $I_{meas}$ were then
obtained from the measured probabilities as in Eq. (\ref{mi}).

\begin{figure}[h!!]
\includegraphics[width=\columnwidth]{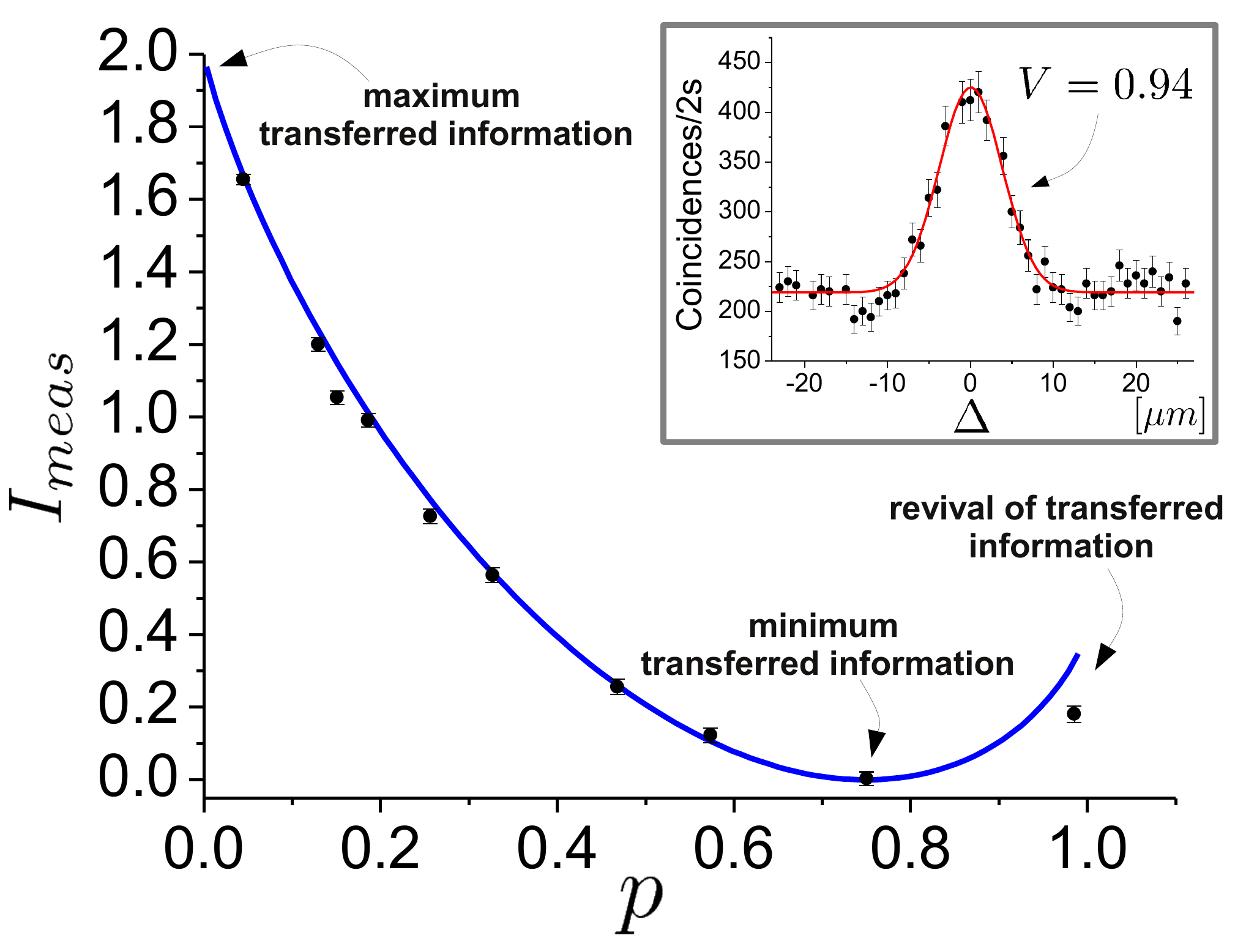}
\caption{Measured values (black dots) of the mutual information and 
theoretical curve (blue line) for the entanglement assisted 
capacity for several values of the parameter $p$ which fully characterize the 
total amount of
depolarizing noise introduced in the experiment. 
The error bars have been obtained by propagating the 
poissonian uncertainties associated to the coincidence counts measured in 10 seconds. 
Inset: peak of the coincidence counts measured in 2 seconds as a function
of the optical path delay $\Delta$ for a state $\ket{\psi^-}\bra{\psi^-}$ entering the 
BS in the absence of controlled noise.}
\label{results}
\end{figure}

The experimental data were analyzed 
by taking into account the non perfect purity of the actual 
input singlet state. This was evaluated for the singlet state $\ket{\psi^-}$ from the visibility 
$\mathcal{V} \approx 94\%$ of the coincidence count peak measured in condition of temporal and spatial
undistinguishability [see the inset of Fig.\ref{results}]. 
According to this value, we could express the
actual input state entering the DC as

\begin{equation}\label{input}
\rho_\mathcal{V}
=\mathcal{V}\ket{\psi^-}\bra{\psi^-} + (1-\mathcal{V})\frac{\id}{4}
\end{equation}

The above state can be interpreted as the result of the action of a preliminar 
DC channel on the pure state $\ket{\psi^-}\bra{\psi^-}$, 
characterized by the noise parameter $p'=3(1-\mathcal{V})/4$.
The action of the sequence of two DCs may be then expressed as a global DC 
with the following global noise parameter
\begin{equation}\label{pnew}
p=\mathcal{V}p_{exp}+p'\;.
\end{equation}

We report in Fig. \ref{results} the experimental results for the transmitted 
information with the theoretical value of the capacity (\ref{midep})
with $p$ given by Eq. (\ref{pnew}). The error bars have been obtained by 
propagating the poissonian uncertainties associated to the coincidence counts.
The agreement between the experimental data
analyzed as explained above and the theoretical behaviour is very high.

{\it Discussions and Conclusions.- }
The theoretical curve in Fig. \ref{results} allows to single out three 
working regimes corresponding to three different values of the noise 
parameter $p$:
\begin{itemize}
 \item $p = 0$ (absence of noise). 
The maximum allowed transferred information consists of 2 bits and the 
noiseless protocol is equivalent to superdense coding \cite{benn92prl}.
 \item $p = 0.75$. The noisy  state is given by the two-qubit maximally 
mixed state $\frac{\id}{4}$. 
In these conditions there is no correlation between the output and the 
input state, hence no information can be transmitted.
 \item $p > 0.75$. In spite of the large amount of noise a partial revival of 
the mutual information is observable due to the non uniform distribution 
of the conditional probabilities $p(y|x)$.
\end{itemize}
The capability of exploring the three regimes strongly depends on the purity 
of the entangled input state. It is remarkable that the  purity, 
attainable from the visibility $\mathcal{V}$ in absence of any
active noise applied by LC3 and LC4 (corresponding to an effective noise 
parameter $p = 0.045$), easily allows to achieve a quite high value of 
$I_{meas}$. Precisely, in this case we
obtained $I_{meas}= 1.655 \pm 0.014$ demonstrating a high channel capacity 
of our system. We want to point out that this value exceeds the threshold for 
linear-optics implementations
reported in \cite{kwiat}, where a superdense coding channel capacity equal 
to $1.630$ bits was obtained. 
Therefore, our result represents also the best performance obtained so far 
in the framework of superdense coding \cite{matt96prl,schae04prl}.
As a further consideration, in these experimental conditions, the growing of mutual information occurring for large 
values of $p$ ($>0.75$) may be clearly detected.

In this work we have given a proof-of-principle experimental demonstration 
of the entanglement assisted capacity for classical information transmission 
over a depolarizing quantum communication channel, 
where classical information is encoded locally on a preshared maximally 
entangled state of two qubits and a controlled noise is then introduced on 
the transmitted qubit. Our experimental implementation of 
the protocol demonstrates the achievement of the classical information 
capacity theoretically predicted for the depolarising channel, therefore
showing the optimal way in which the depolarising channel can be used 
when classical information is to be transmitted and a priori entanglement is 
available.

This work was supported by EU-Project CHISTERA-QUASAR, PRIN 2009 and FIRB-Futuro in ricerca HYTEQ.

\end{document}